\documentclass{iaus}
\usepackage{graphicx}

\title[Comparison of population models] 
{Comparison of different spectral population models}

\author[Koleva et al.]   
{Mina Koleva$^{1,2}$%
  \thanks{mina.koleva@obs.univ-lyon1.fr},
 Philippe Prugniel$^{1,3}$, Pierre Ocvirk$^4$, Damien Le~Borgne$^4$, Igor Chilingarian$^{1,5}$ \and Caroline Soubiran$^6$ }

\affiliation{
$^1$Universit\'e Lyon~1,
Observatoire de Lyon, 9 av. Ch. Andr\'e, St. Genis
Laval, F-69230, France ; CNRS, UMR 5574;\\[\affilskip]
$^2$Department of Astronomy, St. Kl. Ohridski University of Sofia, 5 J.
Bourchier Blvd., BG-1164 Sofia, Bulgaria;\\[\affilskip]
$^3$Observatoire de Paris, GEPI, 61 Ave. de l'Observatoire, Paris, 75014, France;\\[\affilskip]
$^4$CEA Saclay/Service d'Astrophysique
Orme des Merisiers, Bat 709, Gif-sur-Yvette Cedex, F-91191, France;\\[\affilskip]
$^5$Sternberg Astronomical Institute,Moscow University,13 Universitetskij pr., 119992, Moscow, Russia;\\[\affilskip]
$^{6}$Observatoire de Bordeaux,2 rue de l'Observatoire, 33270, Floirac, France;
}

\pubyear{2007}
\volume{241}  
\pagerange{1-2}
\date{30 Jan 2007}
\jname{Proceedings Stellar Populations as Building Blocks of Galaxies}
\editors{A. Vazdekis \& R. Peletier, eds.}
\begin{document}

\maketitle

\begin{abstract}

We have compared simple stellar populations (SSPs)
 generated with different population synthesis tools:
BC03, Vazdekis and P\'egase.HR and different stellar libraries: ELODIE3.1,
SteLib and MILES.
We find that BC03/SteLib SSPs are biased toward solar metallicity, however
P\'egase.HR/ELODIE3.1 and Vazdekis/MILES are extremely consistent.
The extensive coverage of the space of atmospheric parameters in the
large stellar libraries allows precise synthesis for a large range of
ages (0.1 .. 10 Gyr) and
metallicities (-2 .. +0.4 dex) limited by the quality of the determination
of stellar parameters (temperature scale of the giants).

\end{abstract}

\noindent{\bf Introduction.}

There are different methods to derive the parameters of a stellar
population (SSP-equivalent age and metallicity, or population history)
using spectra integrated along the line-of-sight. In this paper we are
using full spectrum fitting (i. e. fitting each pixel).
 We want to estimate the reliability of this approach
and to investigate the consistency between the
different spectral population models and stellar libraries.

Unlike the spectrophotometric indices, this method uses all the
information in the spectrum and is therefore more sensitive (Koleva et al. 2006).
 Though, as well as indices, our approach is
insensitive to the shape of the continuum (plugged with flux calibration
errors and extinction uncertainties). The main ingredients of our method are
the stellar library, the population model and the minimisation procedure.
 
We have tested each of these components 
comparing different stellar libraries  
(SteLib, \cite[Le Borgne 2003]{stelib};
ELODIE3.1, \cite[Prugniel \& Soubiran, 2001, 2004]{elo3};
MILES, \cite[S\'{a}nchez-Bl\'{a}zquez 2006]{Miles}) injected in  
different population synthesis models
(BC03, \cite[Bruzual \& Charlot, 2003]{bc03}; P\'egase.HR, 
\cite [Le Borgne at al. 2004]{peg04};
\cite[Vazdekis 1999]{vaz99}) using different minimisation procedures
(STECKMAP, \cite[Ocvirk et al.]{ocv06} and NBursts, Chilingarian et al., this conference).
As a reference we have used P\'egase.HR evolutionary code with
 ELODIE3.1 (new version to be released soon).  
\\

\begin{figure}
\centering
\resizebox{6.5cm}{!}{\includegraphics{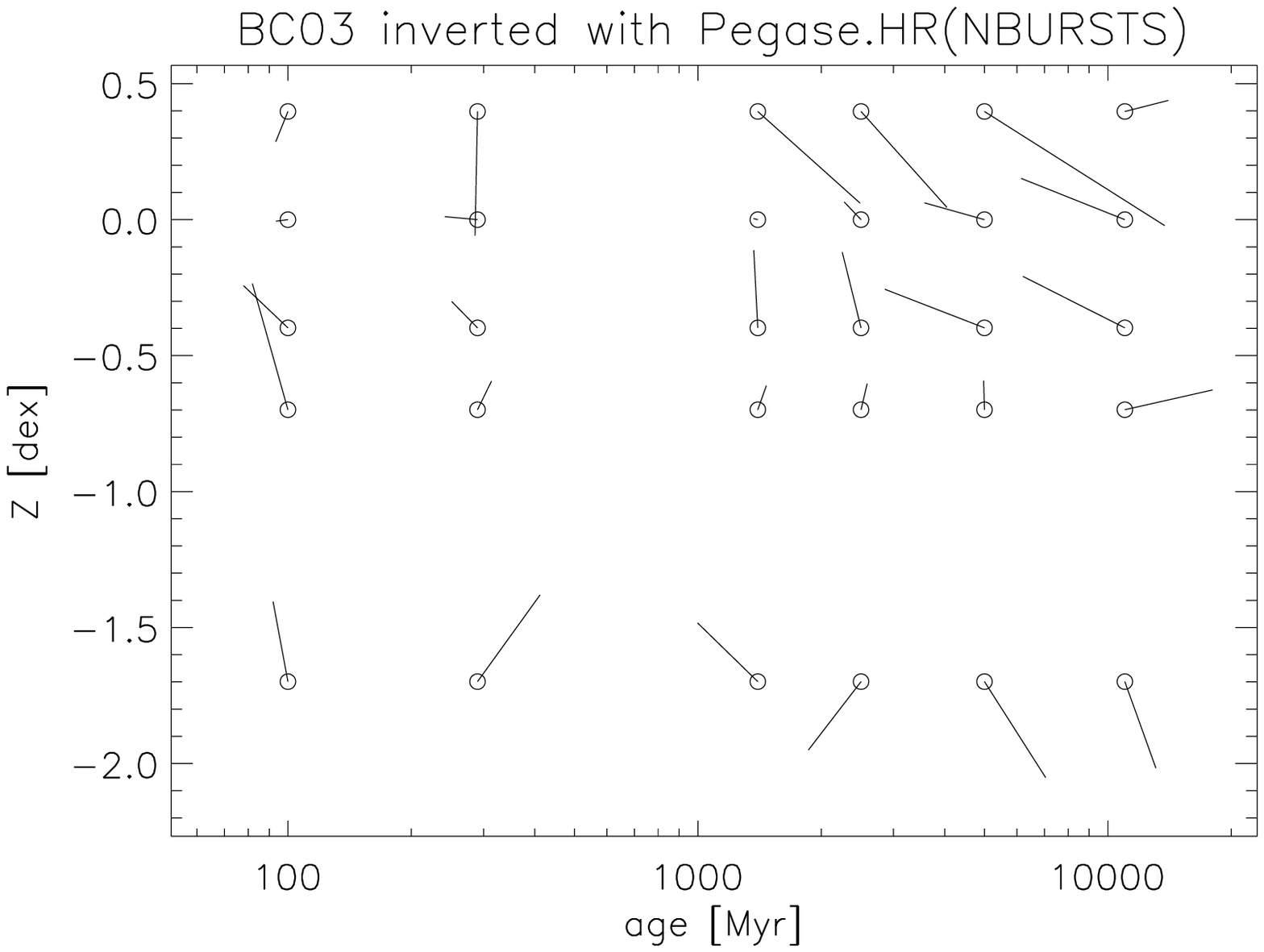} }
\resizebox{6.5cm}{!}{\includegraphics{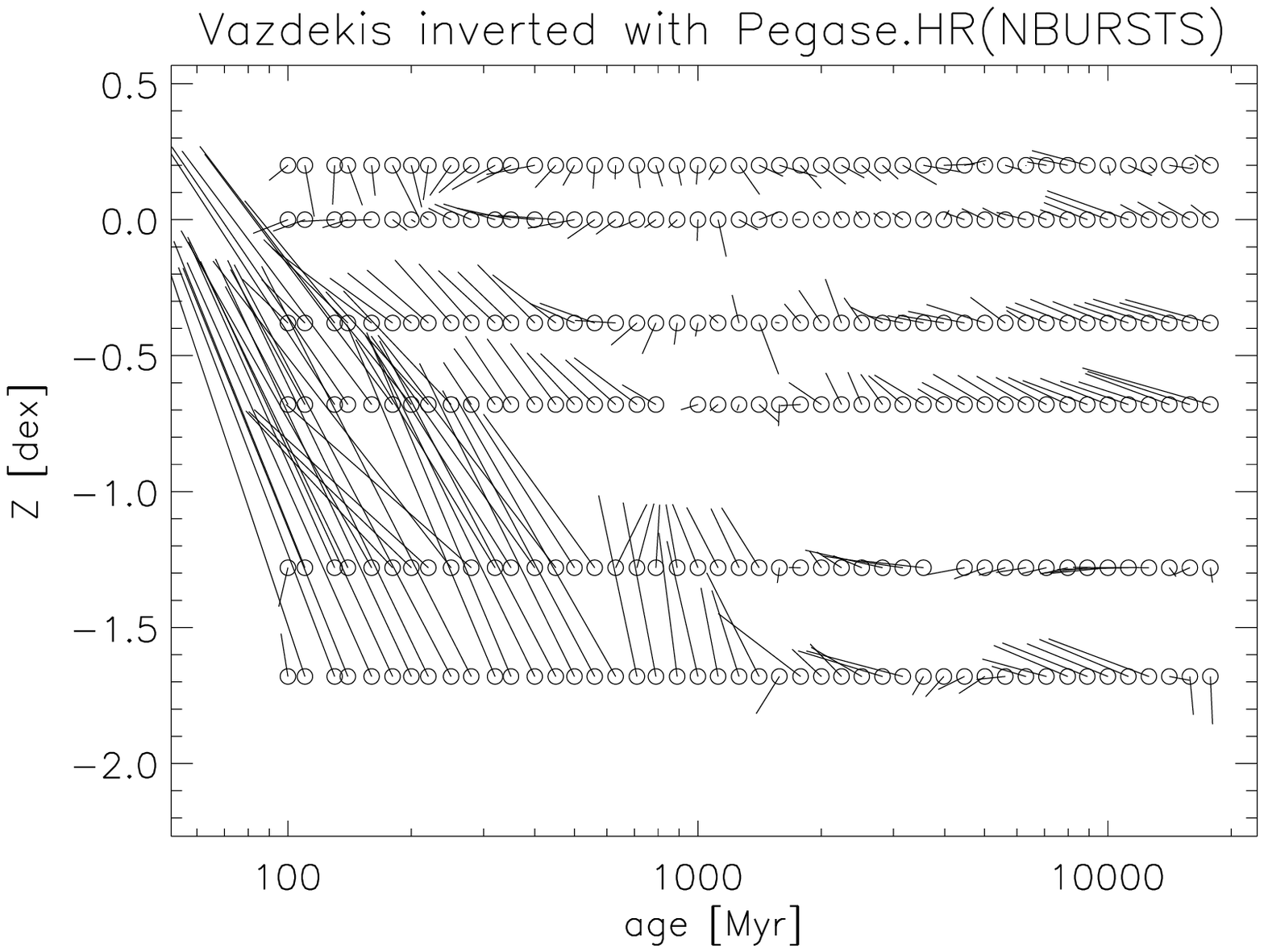} }
\caption {Inversion of (a)Vazdekis-MILES and (b)BC03 models using P\'egase.HR/ELODIE3.1 models. The dots at the location in Age-Metallicity of the BC03 templates are connected to the solution found by inversion.}
\label{Fig1}
\end{figure}

\noindent{\bf Results.}
\begin{enumerate}
\renewcommand{\theenumi}{\arabic{enumi}}
\item     The insensitivity of the fitting algorithm to the
    shape of the continuum produce more robust results;
\item The comparison of BC03 model against P\'egase.HR (Fig. 1a) shows the importance of 
the quality of the libraries (in terms of number of stars and coverage of the 
parametric space). Owing to the small number of stars in SteLib, BCO3 models
 of super- and sub-solar metallicities are using almost solar spectra,
 and the inversion based on more complete libraries (ELODIE3.1 or MILES) converges 
toward solar;
\item The P\'egase.HR and Vazdekis-MILES models (Fig. 1b) are globally consistent,
 and a systematic
 difference between the predicted ages at solar metallicity may be accounted by differences
 in the evolutionary tracks of temperature calibration of the red giants 
(a difference of 200K explains all the differences);
\item The difference between Vazdekis-MILES and P\'egase.HR-ELODIE3.1 metal poor and young SSPs
is related to the lack of hot stars with low metallicity in the stellar libraries. The spectra 
in this part of the diagram are extrapolated;
\item The two used inversion programs STECKMAP and NBursts agree well;
\end{enumerate}
Other validation tests using Galactic clusters are presented in the poster by 
(Koleva et al, this conference).\\

\noindent{\bf Acknowledgements.}
We are grateful to the financial support, provided by the IAU to MK and
IC.

\end{document}